\documentstyle[prl,aps]{revtex}

\newcommand{\vv}[1]{\mbox{\boldmath{$#1$}}}

\def\del{{\rm d}}
\def\beq{\begin{equation}}
\def\eeq{\end{equation}}
\def\sqr#1#2{{\vcenter{\vbox{\hrule height.#2pt
                       \hbox{\vrule width.#2pt height#1pt \kern#1pt
                       \vrule width.#2pt}
                     \hrule height.#2pt}}}}
\def\square{\mathchoice\sqr34\sqr34\sqr{2.1}3\sqr{1.5}3}

\begin{document}  
\twocolumn

\title{Anderson transition of three dimensional phonon modes}

\author{ Yasuyuki  Akita and Tomi  Ohtsuki}
\address{Department of Physics, Sophia University,                       
Kioicho 7-1, Chiyoda-ku, Tokyo 102-8554}

\maketitle

\begin{abstract}
Anderson transition of the phonon modes is studied numerically.
The critical exponent for the divergence of the localization length
is estimated using the transfer matrix method, and
the statistics of the modes is analyzed.
The latter is shown to be in excellent agreement with the
energy level statistics of the disrodered electron system belonging
to the orthogonal universality class.
\end{abstract}


Anderson localization, which was originally proposed for disordered
electron system,\cite{Anderson,KM} is now widely discussed in many
random systems of classical waves, such as photons \cite{photon-loc}
and phonons.\cite{phonon-loc}
Especially interesting is the Anderson transition in three dimensional (3D)
systems, where the wave form changes
from extended to exponentially localized one.
In disordered electron systems, this can be well described by the scaling
theory,\cite{gang,kaw} and universal behaviors of localization length,\cite{KM,slevin-ohtsuki}
conductance distribution \cite{slevin-ohtsuki}
as well as level statistics \cite{level-O,level-U,level-SO} have been demonstrated.
The important point is that the Anderson transition is classified into a few universality
classes.\cite{weg,HLN}

Recently, Song and Kim \cite{song_kim} have shown that the 3D phonon modes
in disordered lattice undergo Anderson transition.
However,
whether this Anderson transition  is classified
into the same universality classes as electron systems remains to be studied. 
In this short note, we demonstrate the Anderson transition of 3D phonon modes
and discuss its critical behavior, and show that it is classified into the orthogonal
universality class of disordered electron system.

We begin with the following equation of motion for the lattice displacement $u(\vv{r},t)$,
\begin{eqnarray}
M\frac{\del^2 }{\del t^2}u(\vv{r},t)
=\sum_{i=x,y,z} &{\ }&
\left[ K_i(\vv{r}-\vv{e}_i) (u(\vv{r}-\vv{e}_i,t)-u(\vv{r},t))+ \right.\\ \nonumber
&{\ }&\left. K_i(\vv{r}) (u(\vv{r}+\vv{e}_i,t)-u(\vv{r},t)) \right]
\end{eqnarray}
where $\vv{r}$ is the lattice position, and
$\vv{e}_i$ the unit vector in the $i$-th direction ($i=x,y,z$).
Here we have assumed the simple cubic structure,
and the displacement is represented by a scalar quantity for simplicity.
$K_i(\vv{r})$ is the spring constant connecting the site at $\vv{r}$
and that at $\vv{r}+\vv{e}_i$.
Only the nearest neighbor interaction is taken into account.

Putting $u(\vv{r},t)={\rm e}^{-{\rm i}\omega t}u(\vv{r})$ and setting $\Omega =\omega^2$,
we reduce the problem to the matrix eigenvalue problem
which is similar to the tight binding model 
where the transfer energy as well as the potential energy is random.
The difference is that off diagonal elements $H_{ij}$'s and the diagonal elements $H_{ii}$'s
are correlated in the phonon system such that
\beq
\sum_j H_{ij}=0 \label{eqn_correl}.
\eeq
%
%
The correlation of the matrix element is very important especially in the
one dimensional systems.
Without the correlation, i.e., when $H_{ii}$'s are independent random
variables, all the states are localized.
When only the off-diagonal elements are random, the states delocalize at the
band center $(E=0)$ with the localization length $\xi\sim |\log E |$ \cite{ER}.
With the above correlation, the localization length diverges according to
the power law of the mode frequency.\cite{ishii}

In the actual simulation, the random spring constant $K_i$'s are
independently and uniformly distributed between $0.1K_0$ and $1.9K_0$.
We scale the frequency $\omega$ by $\sqrt{K_0/M}$ and the length by
the lattice constant.
The density of modes $D(\Omega)$ obtained by diagonalizing $14\times 14\times 14$
systems for 935 configurations is shown in Fig. 1.

We then discuss the critical behavior of the mode $u$.
We apply the transfer matrix method \cite{KM,slevin-ohtsuki} to
calculate the decay length $\lambda_N$ for $N\times N\times L$ system
along $z$-direction, setting $L$ much larger than $N$ and $\lambda_N$.
In actual calculations, $N$ is varied from 8 to 12, and the simulation is
stopped when the relative error of $\lambda_N$ becomes smaller than $2\%$.

In Fig.2, we plot $\Lambda_N =\lambda_N/N$ as the function of $\Omega$.
For extended modes, $\Lambda_N$ is a increasing function with respect to
$N$, while it is a decreasing function for localized modes.
At the critical point $\Lambda_N$ becomes $N$ independent.
The common crossing point in Fig. 2, therefore, gives the critical point
$\Omega_{\rm c}$, which is estimated to be
\beq
\Omega_{\rm c}=13.0\pm 0.2 .
\eeq
The critical exponent $\nu$ for the localization length (or the correlation length) $\xi$
which characterizes the divergence of $\xi$
\beq
\xi\sim |\Omega - \Omega_{\rm c}|^{-\nu}
\eeq
can be estimated by expanding $\Lambda_N$ as
\beq
\Lambda_N= \Lambda_{\rm c} + A_1 N^{1/\nu}(\Omega-\Omega_{\rm c})+\cdots
\eeq
In the present simulation, $\nu$ is estimated to be
\beq
\nu\approx 1.2\pm 0.2   .
\eeq

Recent precise estimate of $\nu_{\rm And}$ in the Anderson model \cite{slevin-ohtsuki}
gives
\beq
\nu_{\rm And}=1.59\pm 0.03
\eeq
for 3D orthogonal universality class, i.e., systems without magnetic fields.
Present estimate of $\nu$ for phonon mode localization length is considerably smaller
than $\nu_{\rm And}$.
This, however, should not be taken seriously, since when the transition occurs near the
band edge, we tend to underestimate $\nu$.\cite{kawara_rf}

In fact, if we study the level statistics regarding $\Omega$ close to the critical point,
we obtain the same critical level statistics obtained for the
Anderson model belonging to the orthogonal universality class.
The distribution function $P(s)$ for the nearest neighbor spacing of the
two successive modes $s$ near the critical points ($12.5<\Omega<13.5$)
is plotted in Fig.3, together with
the critical level statistics for the orthogonal universality class
in electron systems.\cite{level-O}
The coincidence between the two is excellent, and we
conclude that the Anderson
transition of phonon modes is classified into the orthogonal universality class
in the disordered electron systems.
%
%
The critical behavior in 3D is not altered by the correlation of the
matrix elements (Eq.(\ref{eqn_correl})).

In the present model, we have assumed the simple cubic
lattice where the general vector displacement problem
can be reduced to the scalar displacement problem
(isotropic Born model).\cite{feng-sen}
In real lattices, all the components of the displacement vectors
should be taken into account.\cite{feng-sen,YTN}
Such a model is very difficult to analyze by the transfer
matrix calculation.
The present method of level statistics might be a powerful tool
to investigate the phonon Anderson transition
in such  realistic lattices, which is a problem left
for the future.


                                                                                
\medskip                                                                        
The authors are grateful to Professors Yoshiyuki Ono,
Tomoyuki Sekine and
Dr. T. Kawarabayashi for fruitful discussions.
This work is in part financed by the
Grants-in-Aid 09740319
from the Ministry of Education, Science, Sports and Culture. 

\bigskip
%
%

\def\pr{Phys. Rev. }                                                            
\def\prl{Phys. Rev. Lett. }                                                     
\def\jpsj{J. Phys. Soc. Jpn. }                                                  
\def\ssc{Solid State Commun. }

\noindent
{\bf Figure captions}

\medskip

Fig. 1: The density $D(\Omega)$ of the square frequency $\Omega$.

\medskip
 
Fig. 2: Normalized decay length $\Lambda_N$ vs.$\Omega$ for
$N=$8 ($\triangle$), 9($\diamond$), 10 ($\square$) and 12 ($\circ$).
The dashed lines are the 3rd order polynomial fit to the data.
Common crossing point indicates the critical point $\Omega_{\rm c}$, and the
$N$-dependence of the slope at $\Omega_{\rm c}$
gives the localization length exponent $\nu$.

\medskip

Fig. 3: Distribution function $P(s)$ for the nearest neighboring $\Omega$'s close
to $\Omega_{\rm c}$ ($\bullet$).
That for  the disordered electron
systems ($\circ$) 
as well as the Poisson (broken line)
and the Winger (solid line) distributions
is also shown for comparison.

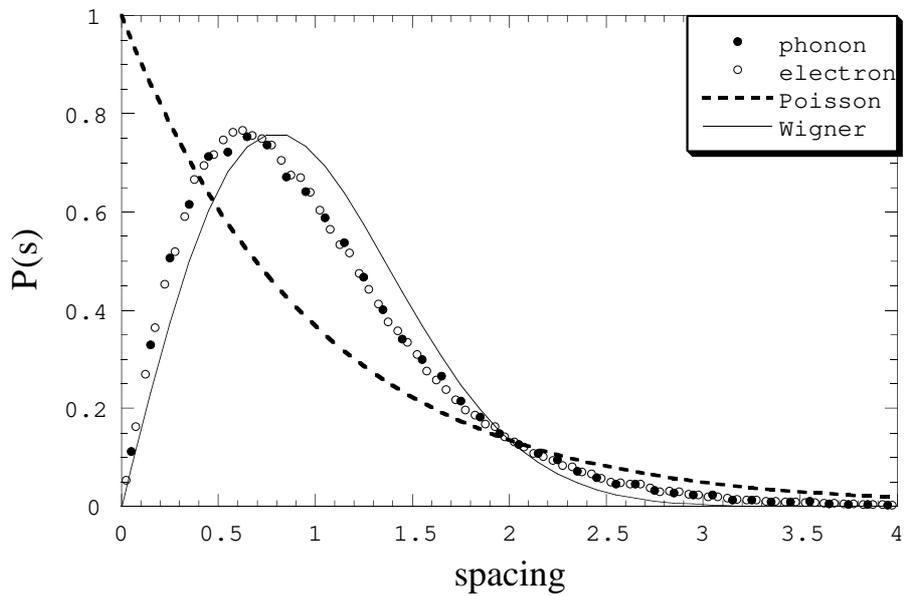
\end{document}